\documentclass[aps, pre, 
amsmath,amssymb,
reprint,%
showpacs,
]{revtex4-1}

\usepackage{graphicx}
\usepackage{dcolumn}
\usepackage{bm}

\usepackage{graphicx}

\newcommand\Rey{\mbox{\textit{Re}}}

\usepackage{color}
\usepackage{transparent}

\newcommand\p{\ensuremath{\partial}}

\begin{document}

\title[Buckling of a beam extruded into highly viscous fluid]{Buckling of a beam extruded into highly viscous fluid}

\author{F. P. Gosselin}%
 \thanks{Corresponding author. Electronic mail: {frederick.gosselin@polymtl.ca.}}
\author{P. Neetzow}
\author{M. Paak}%
\affiliation{ 
Departement of Mechanical Engineering, \'{E}cole Polytechnique de Montr\'{e}al, 2900 Boulevard Edouard-Montpetit, Montr\'{e}al, Qu\'{e}bec, Canada H3T 1J4
}%

\date{\today}

\begin{abstract}

Inspired by microscopic paramecies which use trichocyst extrusion to propel themselves away from thermal aggressions, we propose a macroscopic experiment to study the stability of a slender beam extruded in a highly viscous fluid. Piano wires were  extruded axially at constant speed in a tank filled with corn syrup. The force necessary to extrude the wire was measured to increase linearly at first until the compressive viscous force causes the wire to buckle.
 A numerical model, coupling a lengthening elastica formulation with resistive force theory, predicts a similar behaviour. The model is used to study the dynamics at large time when the beam is highly deformed.
  It is found that at large time, a large deformation regime exists in which the force necessary to extrude the beam at constant speed becomes constant and length-independent. With a proper dimensional analysis, the beam can be shown to buckle at a critical length based on the extrusion speed, the bending rigidity and the dynamic viscosity of the fluid. Hypothesising that the trichocysts of paramecies must be sized to maximise their thrust per unit volume as well as avoid buckling instabilities, we predict that their bending rigidity must be about $3\times 10^{-9}~\mathrm{N\cdot \mu m^2}$. The verification of this prediction is left for future work.

\end{abstract}

\maketitle


\section{\label{sec:Introduction} Introduction}
Self propulsion in micro-organisms is mostly achieved through the beating of flagella or cilia \cite{Bray}.  A widely studied model organism is  unicellular \textit{Paramecium} which swims by beating the cilia covering its body, giving rise to elegant metachronal waves. 
 Although the cell generally employs cilia-based swimming to feed and move, it has been observed recently that when perturbed enough,  paramecium can escape with a ``jumping'' gait \cite{Hamel}.
  To avoid a severe thermal aggression, it can extrude trichocysts in the direction of the aggression to burst away rapidly. These tricocysts are rod-like organelles bound to the plasma membrane. Upon activation, these organelles grow about 10-fold in 3 milliseconds before detaching from the animal \cite{sperling1987crystal,Hamel}.
The drag force created by these extruding rods pushing against the viscous fluid generates thrust in the opposite direction and enables the Paramecium to accelerate itself up to $\sim 10~\mathrm{mm/s}$. 

The function of trichocysts is still a subject of investigation \cite{plattner2002my,buonanno2013defensive}. 
 Their extrusion is known to be a defense mechanism against predators \cite{harumoto1991defensive,buonanno2013defensive}, and it was shown that explosive local release of trichocysts leads to a cell push back as an escape mechanism \cite{knoll1991local}. 
 With their model aggression in the form of local laser heating, Hamel et al. \cite{Hamel} convincingly demonstrated through microscope image time sequences, cell tracking and simple fluid mechanics calculations that paramecium achieves emergency propulsion by trichocyst extrusion.  It comes as a new development in the understanding of micro-organism locomotion \cite{Lauga10052011}, and it raises many questions.

Once fully extended, the trichocysts of paramecium are $40~\mathrm{\mu m}$ in length and just 200~nm in radius.  Considering that trichocysts are left behind after use, they must be regenerated in a process that requires several hours. It is reasonable to expect that these organelles should be optimised to minimise the amount of material that must be left behind to achieve propulsion. Hamel et al. \cite{Hamel} suggest that  trichocysts are well optimised for low Reynolds number viscous flow: ``Because the drag force acting on the trichocysts scales linearly with their length but depends weakly on their width, it is optimal for Paramecium to make the thinnest and longest possible trichocysts.''  From a purely fluid-mechanics point of view, this argument holds true. However, trichocysts are very slender and are loaded in axial compression during deployment.
 Would buckling be a limit to the propulsion burst attainable by trichocyst extrusion? How does buckling occur on an extruding beam?

The extrusion of trichocysts bears resemblance to what is known in dynamics as the inverse spaghetti problem \cite{mansfield1987reverse,paidoussis_fluid-structure_2003}. This problem is encountered in the ejection of paper from a copy machine \cite{mansfield1987reverse,hong2007dynamic},
 morphing aircrafts with deploying wings \cite{zhang2013nonlinear},
 deployment of spacecraft structures such as solar arrays or antennas \cite{downer1993formulation}, extrusion fabrication processes, and robotic manipulators \cite{paidoussis_fluid-structure_2003}. In these different applications, inertia, gravitational, and aero/hydrodynamic forces can come into play.
 Most relevant to the problem at hand is the extrusion of a beam in a dense fluid at large Reynolds number \cite{taleb1981dynamics, gosselin2007stability}. Viscous friction acting against the extrusion can compress the beam and make it unstable, leading to large transverse vibrations. However, in these linear stability studies \cite{taleb1981dynamics, gosselin2007stability}  significant inertia is considered leading to a vibratory dynamics significantly different from the overdamped one expected in the present study at low Reynolds number.
 
Experimental results on the inverse spaghetti problem are few and limited to some work on paper ejection \cite{hong2007dynamic}. Research has been conducted almost exclusively through theoretical models. 
 The inverse spaghetti problem is non-autonomous (i.e., time appears explicitly in the governing equations), and its solution may be sought by time-integration techniques. 
  The dynamics of a deploying Euler-Bernoulli beam has been considered under small displacement assumption \cite{taleb1981dynamics,gosselin2007stability}, with non-linear models coupling axial and transverse deformations \cite{park2013vibrations,zhang2013nonlinear}, and also using elastica models for arbitrary large amplitude \cite{mansfield1987reverse, downer1993formulation,stolte1993extending}. 
 In the studies considering the deployment of a beam in a stagnant fluid \cite{taleb1981dynamics, gosselin2007stability}, the fluid forces on the beam are modelled using a local approach combining an analytically calculated inviscid added mass with semi-empirical pressure drag and skin friction formulations based on the equations of Taylor \cite{taylor1952analysis}.
 More generally, this simplified approach has  been proven succesful for studying the dynamics of beams in large Reynolds number axial flow (\citet[Chapters 8 and 9]{paidoussis_fluid-structure_2003}).
 
 At low Reynolds number, resistive-force theory offers a conceptually similar local drag force model for Stokes flow \cite{gray1955propulsion,Lauga,powers2010dynamics}. In the limit of an asymptotically thin filament, resistive-force theory allows evaluating the hydrodynamic force per unit length acting at a point along the filament independently of the rest of the filament.
 Combined with theoretical modelling, macroscopic scale experimental models of the swimming of micro-organisms have been employed to study the propulsive efficiency of actuated passive rods \cite{Koehler,Kim,Tony,Qian,Coq,rodenborn2013propulsion}. These macro-scale experiments allow elegant demonstration of the physics of micro-organisms as well as a validation of the theoretical fluid-structure interactions models.

In this paper, we present a macro scale experiment inspired by the extrusion of trichocysts by Paramecium. We study the stability of slender steel rods extruded in corn syrup. The high viscosity of the fluid makes the Reynolds number very small and the inertial effects negligible.  We also propose a theoretical model coupling the 2D formulation of a lengthening elastica with the resistive-force theory. This combination of experiments and theory allows understanding how flexibility influences the propulsive force which can be generated by beam extrusion in a viscous fluid as well as investigate its post-buckling dynamics. 


\section{\label{sec:Experiment} Experiment}

\begin{figure}
\includegraphics[scale=1]{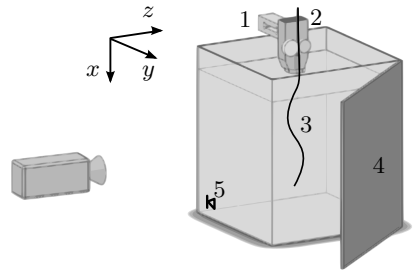}
\caption{\label{fig:tank} Schematic representation of the experimental set-up. The load cell (1) measures the force applied by the feeding motor (2) to extrude the wire (3) in the corn syrup-filled aquarium. Simultaneously the camera records two perpendicular views with a mirror (4). An LED (5) is used to synchronize camera recordings and force data.}
\end{figure}

Steel piano-wires of diameter $d=0.22$~mm and length $L=305$~mm were extruded vertically into a cubic aquarium with sides of $290$~mm filled of highly viscous corn syrup (Fig. \ref{fig:tank}).
 The bending modulus of the wires, $B=22~\mathrm{N~mm^2}$, was determined by hanging weights to their centre and measuring the deflection.
The viscosity of corn syrup was found to be fairly constant $(\mu=4.5~\text{Pa}\cdot\text{s})$ for sheer rates of $2.5$ to $60~\mathrm{s}^{-1}$ in a rheology test conducted with a rotational rheometer. The density of corn syrup was measured to be $\rho=1380~\mathrm{kg~m^{-3}}$.
 The wire is pushed by a  MIG welder wire-drive assembly consisting of a DC motor that drives two counter rotating wheels (USA Weld, Elk Grove Village, IL). 
A 3D printed funnel with a 0.5~mm outlet guides the wire vertically down to the fluid surface. 
 The distance of the outlet of the tip to the motor wheels is approximately $80$~mm, so that the total extrusion length $L_e=225~\mathrm{mm}$. 
 The vertical force, $C_0$, necessary to extrude the wire is measured using a 
\textsc{Tedea-Huntleigh 1022} load cell (Vishay Precision Group, Malvern, PA) 
 connected to a 10-bit \textsc{U3-LV} data-acquisition-system (LabJack, Lakewood, CO). 
  A mirror is placed next to the aquarium so that a camera 
can capture orthogonal views simultaneously.

For each run, the wire is loaded in the feeding mechanism with its tip just above the fluid surface. 
When the power supply is switched on, the wire is extruded at a constant velocity $U$ ranging in equidistant steps from $U = 90~\mathrm{mm/s}$ up to $U=360~\mathrm{mm/s}$. At least 5 tests were conducted for each of the 7 extruding velocity.

Based on the total extruded length, depending on the extruding velocity, the Reynolds number is found to be $\Rey_L={\rho U L_e}/{\mu}\leq 25$. Based on the diameter of the wire, then the value is bound by $\Rey_d={\rho U d}/{\mu}\leq 2.4\times 10^{-2}$. The two corresponding values of Reynolds number in the case of real trichocysts are $\Rey_L=0.52$ and $\Rey_d=2.6\times 10^{-3}$ based on the numbers in Ref. \cite{Hamel}. The Reynolds similitude is satisfactory and we expect the fluid flow in the experiment to be Stokesian.

A representative time sequence of the wire extrusion at a fast speed  is shown in Fig. \ref{fig:beamexp} (a). See Supplemental Material at [URL will be inserted by publisher] for the raw video footage of this extrusion. At first, the wire comes out of the nozzle straight. On the $8^\text{th}$ frame, the wire starts deforming significantly. In the subsequent frames, a distinctive  S-mode of deformation becomes more and more pronounced.
 Note that from inspection of the front and mirror views (not shown), the deformation is largely two-dimensional. 

\begin{figure}
\includegraphics[scale=1]{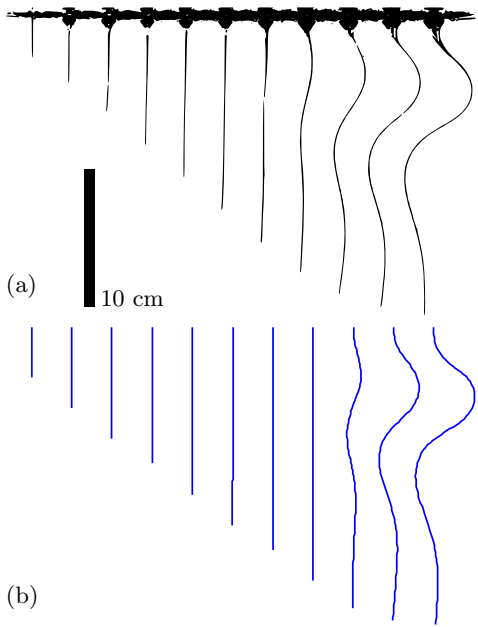}
\caption{Comparison of a time sequence of (a) pictures of a steel piano wire extruded in corn syrup at $U=315\,\text{mm/s}$ with (b) the corresponding modelled deformation. Pictures are taken at $67\,\text{ms}$ intervals. Their background was subtracted and a light intensity threshold was imposed with \textsc{imagej} [National Institutes of Health (NIH), Bethesda, MD]. The modelled deformation at time $tU/\ell=1.6,\,2.5,\,3.4\,..\,10.6$ are obtained for an initial uniform curvature $\kappa_0=0.001$ at $t U/ \ell=1$.}
\label{fig:beamexp}
\end{figure}

\begin{figure}
\includegraphics[scale=1]{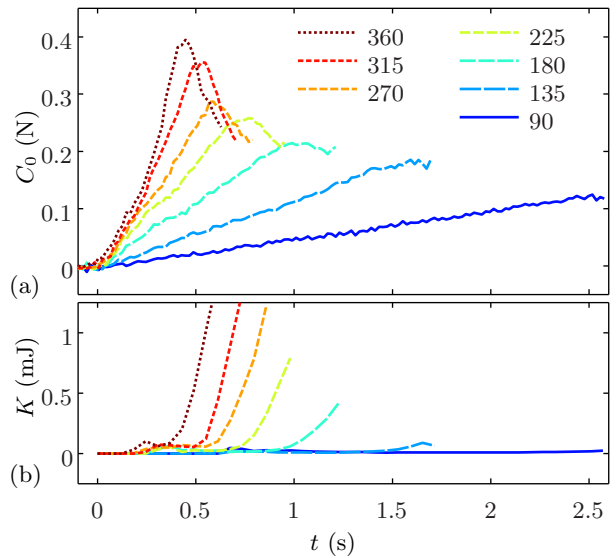}
  \caption{\label{fig:K_F_all} (colour online) Measurements of (a) extrusion force and (b) bending energy as a function of time for 7 different extrusion velocities (measured in mm/s).
  }
  \end{figure}

Force measurements obtained from averaging 5 experimental runs for each extrusion velocity are shown in Fig. \ref{fig:K_F_all} (a). For the two slowest extrusion speeds, the curves display linear increase in the extrusion force with time until the extrusion is stopped. For velocities $180\, \text{mm/s}$ and higher, the extrusion force increases linearly initially until it reaches a maximum value and decreases thereafter. The faster the extrusion speed is, the steeper the initial rise in extrusion force. Also, note that for the fastest extrusion speed $(U=360\text{ mm/s})$, the initial trend does not appear linear because the acceleration of the feeder motor becomes non-negligible. 

From both perpendicular video recordings, image processing in \textsc{Matlab} is used to find two functions describing the instantaneous position of the wire in the $xy$ and $xz$ planes, respectively. It is based on the minimum light intensity for every $x$ value along the vertical. 
  Each of the two generated functions is then least-square fitted on an expansion of 10 cantilevered beam modes. The curvature of these fitted functions in both planes is used to compute the elastic energy of the beam from its clamped end $(x=0)$ to its free end $(x=x_T)$, i.e.,
$ 
K = \frac{1}{2} B\int_0^{x_T}{\left\{ \kappa_y^2+\kappa_z^2 \right\} dx }
$ 
\cite{Landau}. 
The evolution with time of the bending energy computed this way and averaged over 5 experimental runs is shown in Fig. \ref{fig:K_F_all} (b) for the 7 extrusion velocities. For all extrusion velocities, the bending energy is small initially then rises suddenly, indicating a loss of stability by buckling.    
 For faster extrusion speeds, this instability occurs earlier and the rise in bending energy is faster.

To make the considered quantities  dimensionless, we introduce a `bending length' \cite{alben2002drag}, which captures the competition between the wire rigidity and longitudinal viscous fluid forcing:
\begin{equation}
\ell=\left(\frac{B}{ {\zeta}_{\|} U}\right)^{\frac{1}{3}},
\label{eq:characteristic}
\end{equation}
where $\zeta_{\|}$ is the dimensional longitudinal friction coefficient. It is expected that this coefficient vary as $\zeta_{\|}\sim 1/\ln(L/d)$ \cite{powers2010dynamics}, and thus should vary during the extrusion. However, it can be shown that this logarithmic variation has minimal influence on the estimation of the force in the problem at hand \footnote{From slender body hydrodynamics, it is expected that the viscous force on the cylinder obeys a function of the form $f(L)=CL/\log L$. This function can be approximated with a linear fit, which over a range $1\leq L \leq 10$, gives a correlation coefficient $R^2=0.9982$. For our application, an approximation of the viscous force varying linearly with $L$ is accurate enough.}.
  We evaluate the value of $\zeta_{\|}=6.12\,\text{Pa}\cdot\text{s}$ by fitting the relationship  $C_0=\zeta_{\|} U L(t)=\zeta_{\|} U^2 t$ on the force measurement of the slowest extrusion ($U=90\text{ mm/s}$) in Fig. \ref{fig:K_F_all} (a). For this slow extrusion, the wire stays straight and behaves as a stiff cylinder. This one value of the friction coefficient evaluated at $U=90\text{ mm/s}$, is used for calculating the bending length at all extrusion speeds considered here.
 We use the bending length-scale to define an extrusion time-scale $\ell/U$, and a restoring force $B/\ell^2$. Notably, the dimensionless time is equivalent to the instantaneous dimensionless length of the beam $tU/\ell=L/\ell$.

\begin{figure} %
\includegraphics[scale=1]{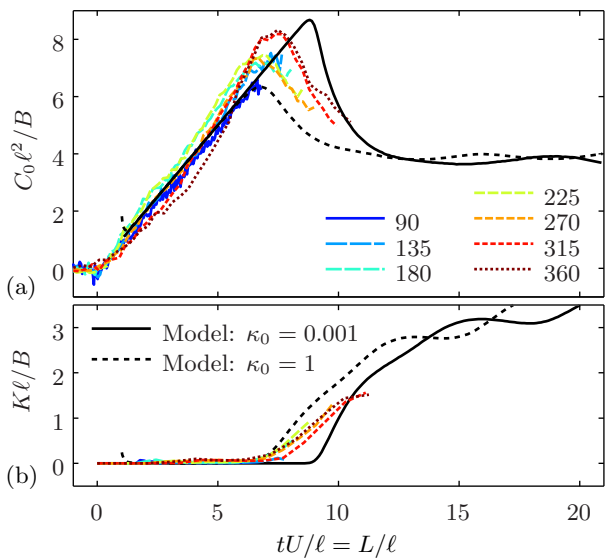}
\caption{\label{fig:dimensionless} (colour online) Dimensionless (a) extrusion force and (b) bending energy measured versus dimensionless time for 7 different extrusion velocities (measured in mm/s); calculations are performed with both a small and a large initial uniform curvature of the beam, $\kappa_0$.}
\end{figure}

The experimental data of Fig. \ref{fig:K_F_all} are plotted in dimensionless form in Fig. \ref{fig:dimensionless}. All the force measurements collapse onto a master curve in Fig. \ref{fig:dimensionless} (a). For time $0<tU/\ell<6$, the dimensionless extrusion force increases linearly, and the bending energy in Fig. \ref{fig:dimensionless} (b) is small indicating that the wire is straight. At dimensionless time between 6 and 8, the extrusion force peaks and the bending energy starts increasing.

\section{\label{sec:Theoretical} Theoretical Model}

For a deeper insight and to consider longer extrusions, we propose a theoretical model coupling a lengtening elastica with a fluid force formulation based on resistive force theory.

 Consider a beam lengthening at a constant speed $U$ such that at any instant $t$, its length is given by $L(t)=Ut$. 
 The deformation of the beam is described by the local angle $\theta(s)$ or curvature $\kappa(s)=\p \theta / \p s$.
 The centreline of the beam is traced by the curvilinear coordinate $s$ defined from its clamped end $(\theta=0\text{, at }s=0)$ to its free end $(\kappa=C=0\text{, at }s=L(t))$. 
 The bending and compressive forces in the beam are balanced by the fluid viscous forces
\begin{equation}
\frac{\p}{\p s}\left( B\frac{\p \kappa}{\p s} \vec{e}_n+C \vec{e}_s \right)=\vec{F},
\label{eq:beam}
\end{equation}
where $B$ is the bending rigidity, $C(s)$ the axial compressive force, $\vec{F}(s)$ the fluid forces acting on the beam, $\vec{e}_s$ the unit tangent vector to the beam, and $\vec{e}_n$ the unit  vector normal to the beam.
 
 For a very slender beam at low Reynolds number, the fluid forces $\vec{F}$ acting on the beam can be approximated using the resistive force theory \cite{gray1955propulsion,Lauga,powers2010dynamics}.  The viscous force per unit length acting at one point on the beam is given by
\begin{equation}
\vec{F}=-\zeta_{\|} v_s \vec{e}_s -\zeta_{\bot} v_n \vec{e}_n, 
\label{eq:fluidforce}
\end{equation}
where $v_s$ and $v_n$ are respectively the local tangential and normal  velocity components of the beam, and where $\zeta_{\|}$ and $\zeta_{\bot}$ are the friction coefficients. For a cylinder in an infinite fluid domain, these friction coefficients can be evaluated using analytical expressions of slender-body hydrodynamics \cite{Keller,gray1955propulsion}. Here, however, we evaluate $\zeta_{\|}$ experimentally  (described below Eq. \ref{eq:characteristic}) and make use of the analytical relationship of $\zeta_{\bot}=2\zeta_{\|}$, true for cylinders of infinite aspect ratio. 


A material point on the beam moves along the centreline at the extruding velocity since we assume the beam to be inextensible, i.e., ${\partial s}/{\partial t} = U$.
With the material derivative, the tangential and normal velocity components can be reconstructed from the Eulerian velocity components
\begin{eqnarray}
v_s =& \frac{\partial x}{\partial t}  \cos \theta + \frac{\partial y}{\partial t} \sin \theta + U,\\
v_n =& -\frac{\partial x}{\partial t}  \sin \theta + \frac{\partial y}{\partial t} \cos \theta,
\end{eqnarray}
where $x$ and $y$ are the eulerian coordinates aligned with the tangential and normal directions when the beam is undeformed.

 Solution of the beam Eq. \ref{eq:beam} with the fluid force of Eq. \ref{eq:fluidforce}  is obtained by integrating the spatial domain with finite differences (function \textsc{bvp4c} in \textsc{Matlab} using default adaptive mesh size).  The Eulerian coordinates are integrated in time using the backwards Euler scheme using time steps of $0.1 \ell/U$.
 The model has only three parameters: 
 we fix $\zeta_{\bot}/\zeta_{\|}=2$, and impose an initial perturbation from the straight equilibrium position in the form of a uniform curvature $\kappa_0$ at time $t U/ \ell$=1.

For a small initial uniform curvature, $\kappa_0=0.001$, the computed shape of the beam as it is extruded is shown in Fig. \ref{fig:beamexp} (b). The similarity with the experimental measurements is appreciable. 

The variation with time of the dimensionless extrusion force and bending energy predicted by the model are shown in Fig. \ref{fig:dimensionless} (a) and (b), respectively. The dynamics of the beam is simulated for a small and a large uniform initial curvature. For $\kappa_0=0.001$, the extrusion force increases linearly at first, then reaches a maximum force of 8.7 at time 8.8 before decreasing and oscillating lightly about a constant value of 3.7 for times larger than 13. The bending energy is small initially and raises abruptly when the extrusion force peaks. For the initial curvature $\kappa_0=1$, the behaviour is the same, but a lower force peak of 6.3 is reached earlier at time 6.8. 
The model agrees well with the experiments: in both, an initial linear force increase is interrupted  when the length reaches a critical value of $L_{cr}/\ell=7\pm 1$. 
 It is informative to compare this prediction of the dynamical non-linear model with linear static theory. 
 A cantilevered beam under a static distributed compressive axial load  (i.e. a standing beam subjected to its own weight) has a critical length $L_{cr}'/\ell=1.986$ beyond which it becomes linearly  unstable \cite{timoshenko1963theory}. This discrepancy is attributable to the fact that upon buckling, perturbations have a finite growth rate and it takes time for the beam to achieve significant deformation when moving in a viscous fluid. Moreover, as the beam keeps lengthening, the bending energy already accumulated gets redistributed in more material.

\begin{figure} 
\centering
\includegraphics[scale=1]{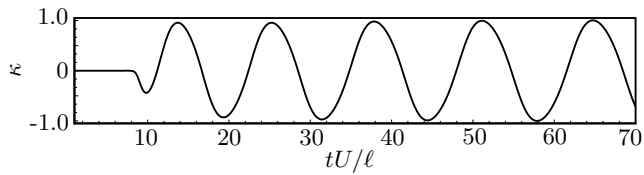}
\caption{\label{fig:base} Variation of the beam dimensionless curvature at the base ($s=0$) versus dimensionless time. The dynamics is computed with the initial uniform curvature $\kappa_0=0.001$ at $t U/ \ell=1$.}
\end{figure}

The non-linear model is used to study the dynamics at large time. Snapshots of the beam shape at incrementing time steps are shown in the inset in Fig. 
\ref{fig:contour}. See Supplemental Material at [URL will be inserted by publisher] for a video animation of this extrusion. At large time, the beam becomes more and more compact. Its deformation shares visual similarities with the inflexional elastica solutions  for the case of a rod subjected to terminal forces alone, without couple \cite{love1944treatise,timoshenko1963theory}.
 The time history of the beam curvature at the base (extrusion point), shown in Fig.~\ref{fig:base}, indicates an oscillatory variation. 
 Similarly to the folding of a falling viscous sheet \cite{PhysRevE.68.036305} or a viscous thread flowing along a less viscous liquid \cite{PhysRevLett.96.114501}, the buckling of a beam extruded in a viscous liquid gives rise to an oscillatory motion. The frequency of oscillation is proportional to the number of bending lengths extruded per second, i.e., $f\propto U / \ell$. Although buckling is usually viewed as a static instability, here it gives rise to a dynamic oscillatory motion.

\begin{figure*} 
\centering
\includegraphics[scale=1]{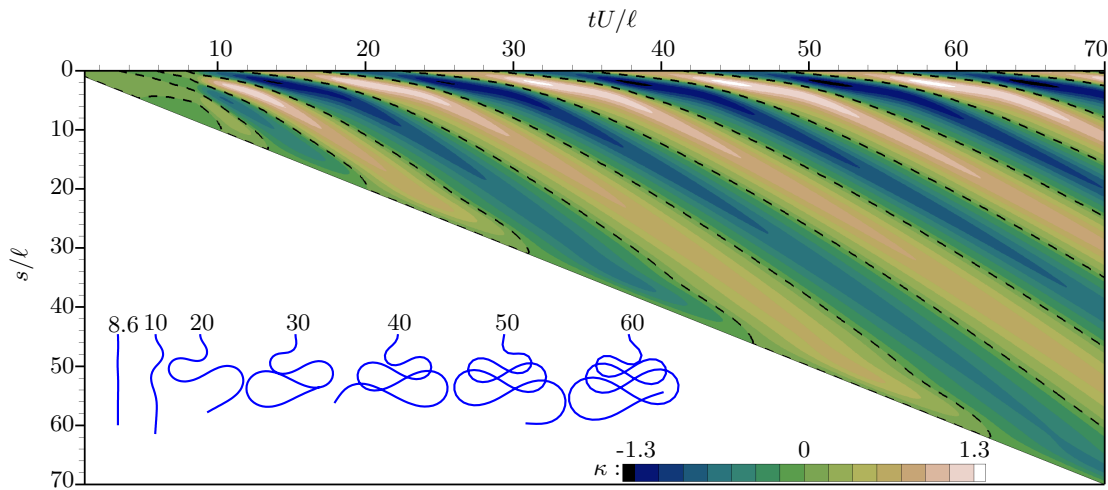}
\caption{\label{fig:contour} (colour online) Contour plot of the beam dimensionless curvature $\kappa$ in the dimensionless space-time plane; dashed lines are the loci of inflection points. The subfigure located inside (blue)  shows the modelled deformation of an extruded beam at different dimensionless time $t U/ \ell$. The dynamics is computed with the initial uniform curvature $\kappa_0=0.001$ at $t U/ \ell=1$.}
\end{figure*}

The contour plot shown in Fig ~\ref{fig:contour} demonstrates the evolution of the beam curvature as the extrusion time and thus the beam length increase.  
In this figure, dashed lines represent the loci of inflection (zero curvature) points, and the regions bounded by these lines are lobes of either positive or negative curvature.
It is seen that, after some initial transient, curvature is generated close to the extrusion point $(0<s/\ell<4)$ and then it moves along the beam as a diminishing-amplitude travelling wave. In other words, after a lobe is generated,  
it moves along the beam while its curvature reduces and finally becomes flat (destroyed) at the beam tip. In terms of speed, the lobes accelerate from a small velocity near the base ($s=0$) to an almost constant velocity for most their journey and then decelerate as approaching the tip.

For times larger than $\sim 13$ and up to the largest simulated time of 70, the extrusion force is constant with only small fluctuations. This means that once the beam is buckled, assuming that its material still behaves elastically, the force required to extrude this forever lengthening beam is constant. This counterintuitive result can be obtained by dimensional analysis following a reasoning similar to \cite{gosselin2010drag}:  at large time, when the beam is  buckled and completely deformed (see inset in Fig. \ref{fig:contour}), its dimension $L$ becomes irrelevant. Thus the only four quantities remaining can be combined into a single dimensionless number required to describe this physical problem at large time, i.e., \smash{$C_0 B^{-1/3}  (\zeta_{\|} U)^{-2/3}$} or equivalently $C_0 \ell^2 B^{-1}$. For a given beam with a fixed bending length and a fixed bending rigidity, the extrusion force must be constant, irrespective of the instantaneous length of the beam.

%
%

It is insightful to compare the behaviour observed in this version of the inverse spaghetti problem at low Reynolds number with the vibrational dynamics simulated in the linear stability analysis of a beam deployed in a dense fluid where inertia is significant 
\cite{taleb1981dynamics,gosselin2007stability}. In both problems, viscous forces can lead to a buckling instability. 
 We have described how this instability influences the extrusion force in the low Reynolds case in Fig. \ref{fig:dimensionless}, however it is not clear how this influence is felt in the problem with inertia because the linear stability analysis employed in these studies \cite{taleb1981dynamics,gosselin2007stability} does not offer this insight. Inertia greatly influences the dynamics of the deploying beam, making it vibrational whereas here the large viscosity leads to an overdamped motion at small time. Interestingly though, an oscillating motion is observed at large time when the extruding beam is highly deformed (see Fig. \ref{fig:base}). Another key difference between both problems is that the model with inertia depends on many different parameters and coefficients, some of which are difficult to evaluate. Here, while it is understood that we made simplifying assumptions regarding the fluid forces, there is basically only the initial conditions which can be varied. And even those do not influence the dynamics tremendously (see Fig. \ref{fig:dimensionless}). When comparing with the complex dynamics of other variants of the inverse spaghetti problem 
\cite{mansfield1987reverse,paidoussis_fluid-structure_2003, hong2007dynamic, zhang2013nonlinear, downer1993formulation, taleb1981dynamics, gosselin2007stability}, the relative simplicity of the dynamics observed for the extruding beam at low Reynolds number is in itself a meaningful result. It is this relative simplicity which allows making predictions about the mechanical nature of trichocysts in the following section.

\section{\label{sec:Trichocyst} Trichocyst Extrusion}

Lastly, we consider how this study applies to trichocyst extrusion. To use material as efficiently as possible because tricocysts must be regenerated after use, we hypothesise that paramecies have evolved trichocysts sized to be as slender as possible yet respecting the critical length of $L_{cr}/\ell=7$. From \cite{Hamel} we can evaluate $L_{cr}=40~\mathrm{\mu m}$,  and $\zeta_{\|}U=17.5\times 10^{-12}~\mathrm{N/\mu m}$. However, to the best of our knowledge, no measurement of the bending rigidity of trichocysts exists in the literature. This makes it impossible to check if trichocysts do respect the critical length. However, if indeed they do, we can predict that their bending rigidity should be about $B=3\times 10^{-9}~\mathrm{N\cdot\mu m^2}$.
To put this number into perspective, the bending rigidity of Paramecium cilia is  $B=6.2\times 10^{-10}~\mathrm{N\cdot\mu m^2}$ \cite{hill2010force}. 
 If our prediction is correct, trichocysts must be 5 times more rigid in bending than cilia.

 Inherent to this calculation of the bending rigidity of trichocysts is the approximation of the fluid forces using resistive force theory where we consider a rod moving in an infinite domain of fluid a rest. 
  It is probable that the presence and the motion of the paramecium cell, and the fact that many trichocysts are extruded in parallel should result in trichocysts perceiving a more complex flow than is considered here. The importance of these effects is hard to estimate without involved CFD analyses. Considering that this resistive force approach was shown to correctly estimate the thrust force of trichocyst extrusion on a swimming paramecium \cite{Hamel}, we can expect it to approximate reasonnably well the internal loads carried by the trichocyst which leads to its instability.
   
\section{\label{sec:Conclusion} Conclusion}

In conclusion, a lengthening elastica model coupled with resistive force theory successfully predicts the time-dependent buckling phenomenon observed when a piano wire is extruded in corn syrup.
 This combination of theory and experiment allowed us to highlight the following aspects of the dynamics of a beam extruded in highly viscous fluid: buckling limits the amount of thrust achievable through beam extrusion; a post-buckling large-deformation regime exists where the force required to extrude a beam at constant speed is constant and  independent of the instantaneous length of the beam; in the post-buckling regime, a self-excited oscillatory motion is generated, sending travelling waves from the root to the tip of the beam;  the bending length is the governing length-scale of the problem which allows collapsing the experimental measurements of the extrusion force and which scales the problem at large time.


Moreover, because we hypothesise that trichocysts must be optimised to be as slender as possible while avoiding large amplitude buckling during their ejection, we predict that their bending rigidity should be about $3\times 10^{-9}~\mathrm{N\cdot \mu m^2}$. Future work should test this prediction, possibly using an atomic force microscope. Confirmation of this prediction would be an additional clue that tricocysts evolved for propulsion. 

\section*{Acknowledgements}
The authors are thankful to S\'ebastien Berger and Ryan Mathews for their design work on preliminary versions of the experimental set-up and to Emmanuel de Langre for enlightening discussions. The financial support of the \emph{Natural Sciences and Engineering Research Council of Canada} and the \emph{Fonds de recherche du Qu\'ebec - Nature et technologies} is acknowledged.

%

\bibliography{aipsamp}

\end{document}